\documentclass[fleqn,a4paper,,draft]{appolb}
\usepackage[intlimits,reqno]{amsmath}
\def\oo{\infty}

\def\NuPh{{\it Nucl. Phys. }}
\def\PL{{\it Phys. Lett. }}

\def\IJMP{{\it Int. J. Mod. Phys. }}

\def\NK{{L}} 
\def\ND{{N_{d}}} 

\def\IOMOG{I^{OM}}
\def\INOMOG{I^{NO}}

\def\e{\epsilon}

\def\K{K}

\def\C{C}
\def\sD{\bar s}

\def\SYS{{\tt SYS}}
\newcommand{\figB}[0]{
\mbox{\parbox{2.6cm}{\hspace{0.4cm}
 \begin{picture}(40,60)(-24,-30)
 \thicklines
   \put(000,000){\circle{40}}
   \put(000,-20){\line(0,+1){40}}
   \put(-20,000){\line(-1,0){10}}
   \put(+20,000){\line(+1,0){10}}
   \put(-040,000){\makebox(0,0)[c]{\footnotesize\text{$p$}}}
   \put(-019,+019){\makebox(0,0)[c]{\footnotesize\text{$m_1$}}}
   \put(-019,-019){\makebox(0,0)[c]{\footnotesize\text{$m_2$}}}
   \put(-007,+000){\makebox(0,0)[c]{\footnotesize\text{$m_3$}}}
   \put(+019,+019){\makebox(0,0)[c]{\footnotesize\text{$m_4$}}}
   \put(+019,-019){\makebox(0,0)[c]{\footnotesize\text{$m_5$}}}
   \end{picture}
}}
\hfill}

\newcommand{\nooa}[1]{
\mbox{\parbox{2.4cm}{\hspace{0.2cm}
 \begin{picture}(30,60)(-32,-33)
 \thicklines
   \put(000,000){\circle{40}}
   \put(000,020){\line(0,1){10}}
   \put(+014.1,014.1){\line(1,1){10}}
   \put(-014.1,014.1){\line(-1,1){10}}
   \put(-018.9,006.6){\line(1,1){12.3}}
   \put(-019.8,002.8){\line(1,1){17.0}}
   \put(-020.0,-000.2){\line(1,1){20.2}}
   \put(-019.8,-002.8){\line(1,1){22.6}}
   \put(-019.3,-005.2){\line(1,1){24.5}}
   \put(-018.6,-007.3){\line(1,1){25.9}}
   \put(-017.7,-009.2){\line(1,1){27.0}}
   \put(-016.7,-011.0){\line(1,1){27.7}}
   \put(-015.5,-012.6){\line(1,1){28.1}}
   \put(-014.1,-014.1){\line(1,1){28.8}}
   \put(-012.6,-015.5){\line(1,1){28.1}}
   \put(-011.0,-016.7){\line(1,1){27.7}}
   \put(-009.2,-017.7){\line(1,1){27.0}}
   \put(-007.3,-018.6){\line(1,1){25.9}}
   \put(-005.2,-019.3){\line(1,1){24.5}}
   \put(-002.8,-019.8){\line(1,1){22.6}}
   \put(-000.2,-020.0){\line(1,1){20.2}}
   \put(002.8,-019.8){\line(1,1){17.0}}
   \put(006.6,-018.9){\line(1,1){12.3}}
   \put(000,-013){\oval(30,40)[b]}
   \put(000,-033){\circle*{4}}
   \put(-000,-038){\makebox(0,0)[c]{\footnotesize\text{$#1$}}}
   \end{picture}
}}
\hfill}
\newcommand{\noob}[1]{
\mbox{\parbox{2.4cm}{\hspace{0.2cm}
 \begin{picture}(30,60)(-32,-33)
 \thicklines
   \put(000,000){\circle{40}}
   \put(000,020){\line(0,1){10}}
   \put(+014.1,014.1){\line(1,1){10}}
   \put(-014.1,014.1){\line(-1,1){10}}
   \put(-018.9,006.6){\line(1,1){12.3}}
   \put(-019.8,002.8){\line(1,1){17.0}}
   \put(-020.0,-000.2){\line(1,1){20.2}}
   \put(-019.8,-002.8){\line(1,1){22.6}}
   \put(-019.3,-005.2){\line(1,1){24.5}}
   \put(-018.6,-007.3){\line(1,1){25.9}}
   \put(-017.7,-009.2){\line(1,1){27.0}}
   \put(-016.7,-011.0){\line(1,1){27.7}}
   \put(-015.5,-012.6){\line(1,1){28.1}}
   \put(-014.1,-014.1){\line(1,1){28.8}}
   \put(-012.6,-015.5){\line(1,1){28.1}}
   \put(-011.0,-016.7){\line(1,1){27.7}}
   \put(-009.2,-017.7){\line(1,1){27.0}}
   \put(-007.3,-018.6){\line(1,1){25.9}}
   \put(-005.2,-019.3){\line(1,1){24.5}}
   \put(-002.8,-019.8){\line(1,1){22.6}}
   \put(-000.2,-020.0){\line(1,1){20.2}}
   \put(002.8,-019.8){\line(1,1){17.0}}
   \put(006.6,-018.9){\line(1,1){12.3}}
   \put(-014.1,-014.1){\line(0,-1){10}}
   \put(+014.1,-014.1){\line(0,-1){10}}
   \put(-014.1,-030){\makebox(0,0)[c]{\footnotesize\text{$#1$}}}
   \put(+014.1,-030){\makebox(0,0)[c]{\footnotesize\text{$#1$}}}
   \end{picture}
}}
\hfill}

\newcommand{\vacuum}[1]{
\mbox{\parbox{0.8cm}{\hspace{0.2cm}
 \begin{picture}(20,30)(-12,-15)
 \thicklines
   \qbezier(0,10)(-12,-8)(0,-10)
   \qbezier(0,10)(+12,-8)(0,-10)
   \put(000,-010){\circle*{4}}
   \put(-000,-016){\makebox(0,0)[c]{\footnotesize\text{$#1$}}}
   \end{picture}
}}
\hfill}
\newcommand{\selfone}[1]{
\mbox{\parbox{1.3cm}{\hspace{0.1cm}
 \begin{picture}(20,30)(-12,-15)
 \thicklines
 \thicklines
   \put(-010,000){\line(-1,0){5}}
   \put(+010,000){\line(1,0){5}}
   \put(000,000){\circle{20}}
   \put(000,-010){\circle*{4}}
   \put(-000,-016){\makebox(0,0)[c]{\footnotesize\text{$#1$}}}
   \end{picture}
}}
\hfill}
\newcommand{\vactwo}[1]{
\mbox{\parbox{1.1cm}{\hspace{0.0cm}
 \begin{picture}(20,30)(-12,-15)
 \thicklines
 \thicklines
   \put(-010,000){\line(1,0){20}}
   \put(000,000){\circle{20}}
   \put(000,-010){\circle*{4}}
   \put(-000,-016){\makebox(0,0)[c]{\footnotesize\text{$#1$}}}
   \end{picture}
}}
\hfill}
\newcommand{\selftwo}[1]{
\mbox{\parbox{1.3cm}{\hspace{0.1cm}
 \begin{picture}(20,30)(-12,-15)
 \thicklines
 \thicklines
   \put(-010,000){\line(-1,0){5}}
   \put(+010,000){\line(1,0){5}}
   \put(-010,000){\line(1,0){20}}
   \put(000,000){\circle{20}}
   \put(000,-010){\circle*{4}}
   \put(-000,-016){\makebox(0,0)[c]{\footnotesize\text{$#1$}}}
   \end{picture}
}}
\hfill}
\newcommand{\selfqa}[1]{
\mbox{\parbox{1.3cm}{\hspace{0.1cm}
 \begin{picture}(20,30)(-12,-15)
 \thicklines
 \thicklines
   \put(-010,000){\line(-1,0){5}}
   \put(+010,000){\line(1,0){5}}
   \put(000,000){\circle{20}}
   \put(7.1,7.1){\circle*{4}}
   \qbezier(-010,000)(-001,001)(000,010)
   \put(+009,014){\makebox(0,0)[c]{\footnotesize\text{$#1$}}}
   \end{picture}
}}
\hfill}
\newcommand{\selfqv}[1]{
\mbox{\parbox{1.3cm}{\hspace{0.1cm}
 \begin{picture}(20,30)(-12,-15)
 \thicklines
 \thicklines
   \put(-010,000){\line(-1,0){5}}
   \put(+010,000){\line(1,0){5}}
   \put(000,000){\circle{20}}
   \put(-7.1,7.1){\circle*{4}}
   \qbezier(-010,000)(-001,001)(000,010)
   \put(-011,014){\makebox(0,0)[c]{\footnotesize\text{$#1$}}}
   \end{picture}
}}
\hfill}
\newcommand{\selfqb}[1]{
\mbox{\parbox{1.3cm}{\hspace{0.1cm}
 \begin{picture}(20,30)(-12,-15)
 \thicklines
 \thicklines
   \put(-010,000){\line(-1,0){5}}
   \put(+010,000){\line(1,0){5}}
   \put(000,000){\circle{20}}
   \put(000,-010){\circle*{4}}
   \qbezier(-010,000)(-001,001)(000,010)
   \put(-000,-016){\makebox(0,0)[c]{\footnotesize\text{$#1$}}}
   \end{picture}
}}
\hfill}
\newcommand{\selfc}[1]{
\mbox{\parbox{1.3cm}{\hspace{0.2cm}
 \begin{picture}(20,30)(-12,-15)
 \thicklines
   \put(000,000){\circle{20}}
   \put(000,-10){\line(0,+1){20}}
   \put(-10,000){\line(-1,0){5}}
   \put(+10,000){\line(+1,0){5}}
   \put(-7.1,7.1){\circle*{4}}
   \put(-011,014){\makebox(0,0)[c]{\footnotesize\text{$#1$}}}
   \end{picture}
}}
\hfill}
\newcommand{\vacuumu}[0]{
\mbox{\parbox{0.8cm}{\hspace{0.2cm}
 \begin{picture}(20,30)(-12,-15)
 \thicklines
   \qbezier(0,10)(-12,-8)(0,-10)
   \qbezier(0,10)(+12,-8)(0,-10)
   \end{picture}
}}
\hfill}
\newcommand{\selfoneu}[0]{
\mbox{\parbox{1.3cm}{\hspace{0.1cm}
 \begin{picture}(20,30)(-12,-15)
 \thicklines
 \thicklines
   \put(-010,000){\line(-1,0){5}}
   \put(+010,000){\line(1,0){5}}
   \put(000,000){\circle{20}}
   \end{picture}
}}
\hfill}

\newcommand\mytoday{\number\day\space \ifcase\month\or
  January\or February\or March\or April\or May\or June\or
    July\or August\or September\or October\or November\or December\fi
      \space\number\year}

\def\eqref#1{Eq.(\ref{#1})}
\def\eqrefb#1#2{Eqs.(\ref{#1})-(\ref{#2})}
\def\itref#1{(\ref{#1})}

\begin{document}
\title{Calculation of Feynman integrals by difference equations\thanks{
Presented at the XXVII International Conference of Theoretical
Physics, {\it Matter to the Deepest: Recent Developments in
Physics of Fundamental Interactions},
Ustro\'n, Poland, 15--21 September 2003.}
\author{S. Laporta\thanks {E-mail: \tt laporta{\char"40}bo.infn.it}
\address{\small \it Dipartimento di Fisica,  Universit\`a di Parma, I-43100
Parma, Italy  \\
Istituto Nazionale di Fisica Nucleare, Sezione di Bologna, I-40126
Bologna, Italy} 
  }
}
\maketitle
\begin{abstract}
In this paper we describe a method of calculation of master integrals
based on the solution of systems of difference equations in one variable.
Various explicit examples are given, as well as the
generalization to arbitrary diagrams. 
\end{abstract}
\section {Introduction}
Calculating Feynman integrals is not easy, especially when diagrams with
many loops are considered. It is by now usual to use 
integration by parts\cite{Tkachov,Tkachov2} identities for reducing 
a given generic Feynman integral to a linear combination of a small number
of irreducible integrals, the so-called ``master integrals''. 
For these an explicit calculation is required. In this paper we illustrate 
a recently developed method of calculation based on the solution of difference
equations. An extended exposition of this method and its applications
can be found in Refs.\cite{Lapdif1,Lapdif2,Lapdif3,Lapdif4}.
We mention here other different approaches with involve
difference equations for calculating one-loop
massive diagrams\cite{Tara1,Tara2}, and structure functions for massless 
diagrams\cite{Verma1,Verma2}.
\section{Difference equations}
Probably the reader is more accustomed to objects like differential equations,
which express relation between functions and their derivatives.
\emph{Difference equations} are a particular kind of functional equations which
express relation between values of functions with argument shifted by
integers. Formally they are identical to ``recurrence relations''  but,
instead of being only a mean to calculate the value of a function from its
neighborhoods, they are functional equation in an unknown function;
in some sense they may be considered as a ``discretized'' form of 
differential equations.
In particular, we will encounter \emph{linear} difference equations.
Difference equation are a well-known mathematical topic 
(G.~Boole $\sim$1850, N.E.~Norlund 1929, L.M.~Milne-Thomson 1933).
A good exposition of the theory and methods of solution can be found 
in the beautiful book~\cite{Milne}.
\subsection{Fibonacci's numbers}
Let us start with a very basic example:
the sequence of Fibonacci's integers $I(n)=\{ 1,1,2,3,5,8,13,21,{\ldots} \}$;
these numbers satisfy the recurrence relation
\begin{equation}\label{fibo}
I(n+2)=I(n+1)+I(n)  \;.
\end{equation}
Now let us think of this relation as a
{\emph{second}-order linear difference equation} 
 with \underline{constants} coefficients.
The general solution of this equation has the form $I(n)=\mu^n$; by
substituting it into the recurrence relation one obtains 
the {\emph{characteristic equation}}
\begin{equation*} 
\mu^2-\mu-1=0  \;,
\end{equation*}
from which we work out the values of $\mu$
\begin{equation*}
\ \mu_1=\dfrac{1+\sqrt{5}}{2} \;,\quad
\mu_2=\dfrac{1-\sqrt{5}}{2}\;.
\end{equation*}
Therefore, the general solution of \eqref{fibo} is an arbitrary 
linear combination of the elementary solutions $\mu_1^n $ and $\mu_2^n$.
In order to reproduce the Fibonacci's sequence of integers we need to fix the
constants, by using the initial conditions $I(1)=I(2)=1$ . One finds 
\begin{equation*}
I(n)=\dfrac{1}{\sqrt{5}} \mu_1^n - \dfrac{1}{\sqrt{5}} \mu_2^n \;.
\end{equation*}
\section {One-loop vacuum integral}
Now let us jump directly to the calculation of the simplest master integral:
the euclidean integral in $D$ dimensions (with unity mass for simplicity)
\begin{equation}\label{inte0}
J(n)=\pi^{-D/2}\int \dfrac{d^D k}{(k^2+1)^n} \equiv \vacuum{n} \;. 
\qquad \qquad \qquad
\qquad \qquad \qquad \qquad \qquad
\end{equation}
By using the identity obtained by integrating by parts\cite{Tkachov, Tkachov2}
\begin{equation*}\label{inte00}
\delta_{\mu\nu}
\int d^D k \dfrac{\partial}{\partial k_\nu} \dfrac{k_\mu}{(k^2+1)^n}= 0 
\end{equation*}
one finds the recurrence relation
\begin{equation}
\label{equ1den}
(n-1)\vacuum{n}-(n-1-D/2)\vacuum{n-1}=0 \;,
\qquad \qquad \qquad \qquad \qquad \qquad \qquad
\end{equation}
which is a \emph{first}-order difference equation 
with \underline{polynomial} coefficients for the function $J$.
A solution of \eqref{equ1den} can be written in the form of 
an \emph{expansion in factorial series}:
\begin{equation}\label{remfac}
\begin{split}
J(n)&=\mu^n \sum_{s=0}^\oo a_s \frac{ \Gamma(n+1)}{\Gamma(n-\K+s+1)}=  \\
&= \mu^n\tfrac{\Gamma(n+1)}{\Gamma(n-\K+1)}\biggl[
         a_0+\tfrac{a_1}{n-\K+1} +\tfrac{a_2}{(n-\K+1)(n-\K+2)}
+\ldots\biggr]\;,
\end{split}
\end{equation}
where $\mu$, $K$, $a_s$ are to be determined.
\subsubsection{Factorial series}
Factorial series play for difference equations the same role
that power series play for differential equations\cite{Milne}.
Factorial series of \emph{first kind} like \eqref{remfac} are similar 
to asymptotic expansions in $1/n$, but with the big advantage of
be convergent! 
For example, let us consider the second derivative of the Gamma function
\begin{equation*}
\Psi'(n)\equiv\dfrac{d^2}{dn^2} \ln \Gamma(n)\;.
\end{equation*}
It satisfies the nonhomogeneous first order difference equation
\begin{equation*}
\Psi'(n+1)-\Psi'(n)=-\dfrac{1}{n^2} \;.
\end{equation*}
If $\Psi'(n)$ is expanded in power series of $1/n$, one obtains 
an \emph{asymptotic} series, not convergent for any value of $n$,
\begin{equation*}
\Psi'(n)= \dfrac{1}{n} + \dfrac{1}{2n^2} +\sum_{k=1}^\oo
\dfrac{B_{2k}}{n^{2k+1}} \;, \qquad
B_{2k}\approx \dfrac{2(2k)!}{(2\pi)^{2k}} \quad \text{if} \ k\to\oo\;,
\end{equation*}
where $B_{2k}$ are the Bernoulli's numbers;
but if one expands it in factorial series, one obtains a 
series comfortably convergent for $n>0$:
\begin{equation*}
\Psi'(n)= \sum_{s=1}^\oo \dfrac{\Gamma(s)}{s}
\dfrac{\Gamma(n)}{\Gamma(n+s)} \;.
\end{equation*}
\subsection{\bf Solution of the difference equation}
By substituting the expansion \itref{remfac} into the difference 
equation \itref{equ1den} (by using the Boole's operators\cite{Milne})
one finds the values $\mu=1$ and $\K=-D/2$, and the recurrence relation
between the coefficients $a_s$
\begin{equation*}\label{recu0}
s a_s= a_{s-1}(s+D/2)(s+D/2-1) \;.
\end{equation*}
The first coefficient $a_0$ is determined by comparing the large-$n$
behaviours of the integral~\itref{inte0} and the factorial 
series~\itref{remfac}:
\begin{align*}\label{inter1}
&\pi^{-D/2}\int \dfrac{ d^D k}{(k^2+1)^n}\approx \pi^{-D/2} \int {d^D k}\; e^{-n k^2}= n^{-D/2}\\
&\mu^n \sum_{s=0}^\oo\frac{a_s \Gamma(n+1)}{\Gamma(n-\K+s+1)} \approx  a_0
 n^\K \mu^n = a_0 n^{-D/2} \qquad 
\Rightarrow \qquad a_0=1  \;.
\end{align*}
For $s\to \oo$  the term of the factorial series behaves as 
\begin{equation*}
\dfrac{a_s}{\Gamma(n+D/2+s+1)} \approx  \dfrac{s!\ s^{D-1}}{s! \ s^{D/2+n}}
=  s^{-1+D/2-n}\;;
\end{equation*}
$n=D/2$ is the \emph{abscissa of convergence} of the factorial series, in fact:
\begin{itemize}
\item if $n>D/2$ the series is convergent, and it converges quickly for
      $n \gg 1$;
\item if $n \le D/2$ the series does not converge, but 
      we can calculate $J(n+i)$ for some large integer $i$ by using the 
      series, and then to use repeatedly the
      recurrence relation \itref{equ1den} in order to obtain 
      $J(n+i-1)$, $J(n+i-2)$, ${\ldots}$, $J(n)$.
\end{itemize}

\section{On-mass-shell self-mass master integral}
\begin{equation*}\label{inte1}
I(n)=\pi^{-D/2}\int \dfrac{d^D k}{(k^2+1)^n(k^2-2 p\cdot k)}
\equiv\selfone{n} \;.
\end{equation*}
By combining identities obtained integrating by parts 
one finds 
a \emph{nonhomogeneous} \emph{second} order difference equation
\begin{equation}\label{diffequ1a}
( n - D )\selfone{n-2}+( 2n - D - 1 )\selfone{n-1} -3(n-1)\selfone{n}=
(1 -\tfrac{D}{2})\vacuum{n - 1} \;.
\end{equation}
The general solution of this equation has the form
\begin{equation*}\label{solgeni}
I(n)=\C_{1}I_{1}(n) + \C_{2}I_{2}(n) +I_3(n)\;,
\end{equation*}
where $I_1$ and $I_2$ are solutions of the homogeneous equation, 
$\C_1$ and $\C_2$ arbitrary constants,
and $I_3$ is a particular solution of the nonhomogeneous equation.
As before, we look for solutions in the form of expansions in factorial series
\begin{equation}\label{enpfact}
I_j(n)=\mu_j^n 
    \sum_{s=0}^\oo\frac{b_s^{(j)} \Gamma(n+1)}{\Gamma(n-\K_j+s+1)}\;, \quad 
j=1,2,3\;.
\end{equation}
By substituting \eqref{enpfact} into
\eqref{diffequ1a} one finds the values
\begin{alignat*}{3}
\mu_1=&1,       \qquad&\mu_2=&-1/3,      \qquad& \mu_3=&1, \\
\K_1=&-D/2+1/2, \qquad& \K_2=&-D/2+1/2 , \qquad&  \K_3=&-D/2\;,
\end{alignat*}
and the recurrence relations 
\begin{equation}\label{recu1}
4s b_s^{(1)} -(\sD-\tfrac{1}{2})(7\sD-D+\tfrac{1}{2})b_{s-1}^{(1)}+3(\sD
-\tfrac{3}{2})(\sD-\tfrac{1}{2})^2 b_{s-2}^{(1)}=0\;,
\end{equation}
\begin{equation}\label{recu2}
(4s+2)b_s^{(3)} -\sD(7\sD-D+4)b_{s-1}^{(3)} 
+3\sD^2(\sD-1)b_{s-2}^{(3)}= (\tfrac{D}{2}-1)(a_s-\sD a_{s-1}) \;,
\end{equation}
where $\sD = s+\tfrac{D}{2}-1$, $a_i=b_i=0$ for $i<0$ and  
$b_0^{(1)}=b_0^{(2)}=1$. 

The constants $C_1$ and $C_2$ are determined
by comparing the behaviours for $n\to\oo$, 
$I_1(n)\approx n^{-D/2+1/2}$, $I_2(n)\approx (-1/3)^n n^{-D/2+1/2}$ and
$I_3(n)\approx (1/2-D/4) n^{-D/2}$ with 
\begin{equation*}
\selfone{n}\approx \pi^{-D/2} \int  \dfrac{d^D k\; e^{-n k^2}}{k^2-2p \cdot k} 
\approx\dfrac{\sqrt{\pi}}{2}  n^{-D/2+1/2}\;;  
\qquad \qquad \qquad \qquad \qquad
\end{equation*}
one finds $\C_1=\tfrac{\sqrt{\pi}}{2}$, $\C_2=0$, therefore the solution $I_2$
can be discarded.
As in the previous case, the factorial series converge quickly
for $n \gg 1$, and the recurrence relations are used for the values of $n$
      such that the series converges slowly or does not converge.
\section{General case}
Now we sketch the generalization of the method used above to the 
calculation of master integrals of arbitrary diagrams.
Given an $L$-loop integral,
we put the exponent $n$ on the first denominator (arbitrarily chosen)
\begin{equation*}
B(n)=
\pi^{-DL/2}\int 
{d^D k_1 \dots d^Dk_L}
\dfrac{N}
{D_{1}^n D_{2} D_{3} \ldots D_{\ND}} \ ; 
\end{equation*}
the numerator $N$ in general contain products of powers of some irreducible
scalar products involving loop momenta.
By combining by suitable algorithms the identities obtained by
integration by parts 
\begin{equation*}
\delta_{\mu\nu}
\int d^D k_1 {\ldots} d^Dk_\NK 
\dfrac{\partial}{\partial k_{i\nu}} \dfrac{\{k_{1\mu} ; k_{2\mu} ; {\ldots};
p_{1\mu}; {\ldots}\}\ N_j} 
{D_1^n D_2^{\alpha_2} D_3^{\alpha_3}  {\ldots} D_\NK^{\alpha_\NK}}= 0 
\end{equation*}
one finds a nonhomogeneous difference equation of order $r$
(typically $r\sim2-10$)
\begin{equation*}\label{diffu1}
p_0(n)B(n) +p_1(n)B(n+1)+{\ldots} +p_r(n)B(n+r)=T(n)\;;
\end{equation*}
the right-hand side $T(n)$ contains new integrals $B_1$, $B_2$,{\ldots}
which have one or more denominators missing with respect to  $B$, 
as an example
\begin{equation*}
B_1(n)=
\pi^{-DL/2}\int 
{d^D k_1 \dots d^Dk_L}
\dfrac
{N}
{D_{1}^n D_{3} D_{4} \ldots D_{\ND}} \;,
\end{equation*}
and which are simpler.
By building difference equations for $B_1$, $B_2$, {\ldots}  
one obtains a system of difference equations in triangular form.
The system is solved one equation at once, 
beginning with that corresponding to the simplest master integral,
with the minimum number of denominators.
The solutions 
\begin{equation*}
B_l(n)= \C_1 \IOMOG_1(n) + \C_2 \IOMOG_2(n) + {\ldots} \C_r \IOMOG_r(n) +
 \INOMOG(n)  
\end{equation*}
are expanded in series 
\begin{align*}\label{remfac2}
\IOMOG_j(n)&=\mu_j^n \sum_{s=0}^\oo\frac{a_s^{(j)} 
\Gamma(n+1)}{\Gamma(n-\K_j+s+1)}\;, \\
\quad
\INOMOG(n)&=\sum_{j=1}^S (\mu^{NO}_j)^n \sum_{s=0}^\oo\frac{a_s^{(j,NO)} 
\Gamma(n+1)}{\Gamma(n-\K_j^{NO}+s+1)}\;;
\end{align*}
$\mu_j^{NO}$, $K^{NO}_j$ and $a_s^{(j,NO)}$ are obtained from the
expansion of $T(n)$ which is supposed to be known.
By substituting the expansions into the difference equation one finds
the characteristic equation and the values of  $\mu_1, \mu_2,{\ldots}, \mu_r$, 
the $r$ indicial equations  and the values of  $K_1$,{\ldots},  $K_r$,
and the recurrence relations for the coefficients $a_s^{(j)}$ 
which have usually order higher than \eqrefb{recu1}{recu2}($\sim$10-20). 
By comparing the behaviours for $n \to \oo $ the values
of the constants $\C_1$, $\C_2$, {\ldots}, $\C_r$ are obtained (usually 
most of constants are null).
For example, the large-$n$ behaviour of an euclidean massive integral is: 
\begin{equation*}
B(n)\approx \left(\dfrac{1}{m_1^2}\right)^{n-D/2} n^{-D/2} F(0) \;,
\end{equation*}
where $D_1=k_1^2+m_1^2$, $m_1\not = 0$.
$F(0)$ is the value of the integral corresponding to the diagram 
with one loop less obtained  by setting $k_1$=0, graphically
\begin{equation*}
\nooa{n} \approx \left(\dfrac{1}{m_1^2}\right)^{n-D/2} n^{-D/2}
\noob{\smash{k_1=0}} \;.
\end{equation*}
\section{System of difference equations in the single-scale case}
To give to the reader an idea of how the difference equations for
more complicated diagrams look, we list here the whole system of difference 
equations for the two-loop self-mass diagram in the particular case
with all masses equal to $1$ and on-shell external momentum.
As above, the point on a line means that the corresponding denominator is raised
to the power indicated. The factorizable two-loop subdiagrams are decomposed 
into the product of one-loop diagrams.
\begin{multline*}
\hspace{-7mm}
(n-1)\vacuum{n}-(n-1-D/2)\vacuum{n-1}=0 \;,
\qquad \qquad \qquad \qquad \qquad \qquad \qquad 
\end{multline*}
\begin{multline*}
\hspace{-7mm}
( n - D )\selfone{n-2}+( 2n - D - 1 )\selfone{n-1} -3(n-1)\selfone{n}=
 (1 -\tfrac{D}{2})\vacuum{n - 1} \;,
\quad 
\end{multline*}
\begin{multline*}
\hspace{-7mm}
( n - D ) \vactwo{n-2} +(2n-D-1)\vactwo{n-1}-3(n-1)\vactwo{n}=
  (2-D)\vacuum{n-1}\times\vacuumu\;,
\end{multline*}
\begin{multline*}
\hspace{-7mm}
    (n-D)(2n-3D+2)\selftwo{n-2}
    +2[7n^2-n(10D+1)+3D^2+3D-2]\selftwo{n-1}\\
\qquad \qquad \qquad 
	    -16(n-1)(n-D+1)\selftwo{n} = 
	    2(D-2)^2\vacuum{n-1}\times\vacuumu\;,
\end{multline*}
\hspace{-7mm}
\begin{multline*}
        (2n-3D+4)\selfqa{n-2}
	+(4n-D-2)\selfqa{n-1}
	-6(n-1)\selfqa{n}= \qquad \qquad \\
\quad \qquad \qquad \qquad \qquad 
	-(D-2)\vactwo{n-1}-2(D-2)\selfone{n-1}\times\vacuumu\;,
\end{multline*}
\begin{multline*}
\hspace{-7mm}
48(n-D+1)\biggl[
(n-D)\selfqb{n-2}       
+(2n-D-1)\selfqb{n-1}	    
-3(n-1)\selfqb{n}\biggr]=
\\
\qquad
-5(n-D)(2n-3D+2)\selftwo{n-2}
-[- 10n^2+n(28D-58)-18D^2\\
\qquad
+ 46D+12]\selftwo{n-1} 
-(D-2)(-32n+6D+52)\vacuum{n-1}\times\vacuumu\;,
\end{multline*}
\begin{multline*}
\hspace{-7mm}
48(n-D+1)\biggl[
(n-D)\selfqv{n-2}       
+(2n-D-1)\selfqv{n-1}	    
-3(n-1)\selfqv{n}\biggr]=   
\\
-5(n-D)(2n-3D+2)\selftwo{n-2}
-[- 10n^2+n(28D-58)-18D^2+ 46D\\+12]\selftwo{n-1}
-24(D-2)(n-D+1)\vacuum{n-1}\times\selfoneu 
\qquad \qquad \qquad \qquad
\\
-2(D-2)(8n-9D+2)\vacuum{n-1}\times\vacuumu\;,
\qquad \qquad \qquad
\qquad \qquad \qquad \qquad
\end{multline*}
\begin{multline*}
\hspace{-7mm}
36(n-D+1)\biggl[
 2(n-D+1) \selfc{n-2} 
 +(2n-D)\selfc{n-1}
 -4(n-1)\selfc{n} \biggr]= \\
+12(n-D+1)\biggl[ 
       (2n-3D+4)\selfqa{n-2} -2(n-D+1)\selfqa{n-1} \\
       \qquad \qquad 
       -(n-D)\selfqb{n-2} 
       +(n-2D+4)\selfqb{n-1} 
       -(n-D)\selfqv{n-2}  \\
 \qquad \qquad
 \qquad
       +(n-2D+4) \selfqv{n-1}
       \biggr] 
       -(n-D)(2n-3D+2)\selftwo{n-2} \\
       -(- 2n^2  - 16nD + 46n + 18D^2 - 70D  +60)\selftwo{n-1}
 +12(n-D+1) \biggl[ \\
 \qquad \qquad
   -(n-D)\selfone{n-2}\times\selfoneu 
   +(n-2D+4)\selfone{n-1}\times\selfoneu \\
 \qquad
 +(D-2)\vactwo{n-1} 
 -(D-2)\vacuum{n-1}\times\selfoneu -(D-2)\selfone{n-1}\times\vacuumu
 \biggr] \\
 \qquad \qquad \qquad
 \qquad \qquad \qquad
     +2(D-2)(8n-3D-10)\vacuum{n-1}\times\vacuumu\;.
\end{multline*}

\section{\bf The program \SYS }

A generic integral, through the solution of a  
system of  algebraic identities between integrals
obtained integrating by parts,
is transformed into 
a linear combination of a small number of  
``master integrals'',  chosen typically with numerator unity  and
denominators raised to power one.

The method of calculation based on the solution of finite difference 
equations is applied only to these ``master integrals''.

Considered the number and the complexity 
of the systems of difference equations for diagrams with more than  2 loops,
we have written on purpose a comprehensive program, called \SYS, with the
following features:
\begin{itemize}
\item C program, about 23000 lines.
\item The program automatically determines the master integrals of a diagram,
      it builds and solves the systems of difference equations.
\item Input: description of the diagram, 
      number of terms of the expansion in $D-4$.
\item The program contains a simplified algebraic manipulator,
      used to solve systems of identities among integrals with 
      this kind of coefficients:
      arbitrary precision integers, rationals,
      ratios of polynomials in one and two variables
      (for example $D$ and $x$) with integer coefficients.
\item Efficient management of systems of identities of size up to
      the limit of disk space (tested up to $5\cdot10^6$ identities).
\item Numerical solution of systems of difference and differential equations
      up to 500 equations, using  arbitrary precision floating point complex
      numbers and truncated series in $\e$.
\item All the coefficients of the expansions in $\e$ 
      are worked out in numerical form, even those of divergent terms.
\item Floating number precision: up to 1200 digits (see \cite{Lapdif4})
      (essentially one sums expansions in  \emph{one} variable).
\item Arithmetic libraries which deal with operations on integers,
      polynomials, rationals, floating point numbers and truncated
      series in $\e$ were written on purpose by the author.
\end{itemize}

\section{\bf Applications of the method}
\begin{enumerate}
\item high-precision calculation of vacuum (2-4 loops), self-mass
(1-3 loops), vertex and box diagrams (1-2 loops) in the case of 
equal masses and on-shell external momenta \cite{Lapdif1,Lapdif2,Lapdif4}.
\item high-precision calculation of electron $g-2$ master integrals,
     at 3 loops \cite{Lapdif3} and at 4 loops\cite{sun4}
     (actually only a few, work in progress).
\item high-precision calculation of integrals for \emph{arbitrary}
      values of mass and momenta (work in progress).
\end{enumerate}
Here is an example of high-precision (60 digits) result obtained
with the program \SYS: the expansion in 
$\e=(4-D)/2$ of the following two-loop self-mass integral with these values of 
masses and momenta:
$p^2=-1$, $m_1^2=2$, $m_2^2=3$, $m_3^2=4$, $m_4^2=5$, $m_5^2=6$,
{
\footnotesize
\begin{multline*}
\hspace{-1cm} \figB =\Gamma(1+\e)^{2}\bigl[
\qquad\qquad\qquad\qquad
\qquad\qquad\qquad\qquad
\qquad\qquad\qquad\qquad\quad
\\
\hspace{-4pt}
+0.213887190522735619987564582191165767549812066097826018573\\
\hspace{-4pt}
-0.918487358981510992237303382863013109752582619839173727799\e\\
\hspace{-4pt}
+2.395114573687568040190458275515297803861359807588836500079\e^2\\
\hspace{-4pt}
-5.204471820133273025267916826530760950863731824065596549403\e^3\\
\hspace{-4pt}
+10.62755429637642806105390035038189357736980511197603387938\e^4
+O(\e^5)\bigr] \;.
 \end{multline*}
 }
\section{Conclusions}
\begin{enumerate} 
\item Method of calculation 
      applicable to diagrams with any topology.
\item The calculation of an integral is reduced to the sum of convergent
      series in \emph{one} variable.
      As consequence, the number of operations needed to obtain a precision
      of the results of $N$ digits grows linearly with $N$ and
      \emph{does not depends on the complexity of the diagram};
      calculations with $N\sim10^3$ are still feasible\cite{Lapdif4}.
\item Easiness of cross-checks, due to possibility of writing  
      different difference equations for each topologically 
      distinct line of a given diagram.
\end{enumerate}


\end{document}